\documentclass[aps,prb,twocolumn,reprint,superscriptaddress,floatfix,citeautoscript,longbibliography]{revtex4-1}
\usepackage[T1]{fontenc} 
\usepackage[utf8]{inputenc}
\usepackage{lmodern}
\usepackage{graphicx}
\usepackage{amsmath}
\usepackage{siunitx}
\usepackage[colorlinks=true,linkcolor=blue,citecolor=blue,filecolor=blue,urlcolor=blue]{hyperref}

\newcommand{\nmr}{NMR}

\newcommand{\bnmr}{$\beta$-\nmr}

\newcommand{\musr}{$\mu$SR}
\newcommand{\epr}{EPR}
\newcommand{\Endor}{ENDOR}
\newcommand{\slr}{SLR}
\newcommand{\efg}{EFG}
\newcommand{\rf}{RF}
\newcommand{\cw}{CW}

\newcommand{\eli}{\textsuperscript{8}{Li}}
\newcommand{\elip}{\textsuperscript{8}{Li}\textsuperscript{+}}

\newcommand{\lip}{{Li}\textsuperscript{+}}

\newcommand{\ROOT}{ROOT} 
\newcommand{\minuit}{MINUIT}

\newcommand{\srim}{SRIM}

\newcommand{\rutile}{rutile TiO$_{2}$}

\begin{document} 

\title{Microscopic Dynamics of \lip\ in Rutile TiO$_2$ Revealed by \eli\ \texorpdfstring{$\beta$}{b}-detected NMR}

\author{Ryan M. L. McFadden}
\email[]{rmlm@chem.ubc.ca}
\affiliation{Department of Chemistry, University of British Columbia, 2036 Main Mall, Vancouver, BC V6T~1Z1, Canada}
\affiliation{Stewart Blusson Quantum Matter Institute, University of British Columbia, 2355 East Mall, Vancouver, BC V6T~1Z4, Canada}

\author{Terry J. Buck}
\affiliation{Department of Physics and Astronomy, University of British Columbia, 6224 Agricultural Road, Vancouver, BC V6T~1Z1, Canada}

\author{Aris Chatzichristos}
\affiliation{Stewart Blusson Quantum Matter Institute, University of British Columbia, 2355 East Mall, Vancouver, BC V6T~1Z4, Canada}
\affiliation{Department of Physics and Astronomy, University of British Columbia, 6224 Agricultural Road, Vancouver, BC V6T~1Z1, Canada}

\author{Chia-Chin Chen}
\affiliation{Max-Planck-Institut f{\"u}r Festk{\"o}rperforschung, Heisenbergstra{\ss}e 1, 70569 Stuttgart, Germany}

\author{David L. Cortie}
\altaffiliation{Current address: Research School of Chemistry, Australian National University, Canberra, ACT 2601, Australia}
\affiliation{Department of Chemistry, University of British Columbia, 2036 Main Mall, Vancouver, BC V6T~1Z1, Canada}
\affiliation{Stewart Blusson Quantum Matter Institute, University of British Columbia, 2355 East Mall, Vancouver, BC V6T~1Z4, Canada}
\affiliation{Department of Physics and Astronomy, University of British Columbia, 6224 Agricultural Road, Vancouver, BC V6T~1Z1, Canada}

\author{Kim H. Chow}
\affiliation{Department of Physics, University of Alberta, 4-181 CCIS, Edmonton, AB T6G~2E1, Canada}

\author{Martin H. Dehn}
\affiliation{Stewart Blusson Quantum Matter Institute, University of British Columbia, 2355 East Mall, Vancouver, BC V6T~1Z4, Canada}
\affiliation{Department of Physics and Astronomy, University of British Columbia, 6224 Agricultural Road, Vancouver, BC V6T~1Z1, Canada}

\author{Victoria L. Karner}
\affiliation{Department of Chemistry, University of British Columbia, 2036 Main Mall, Vancouver, BC V6T~1Z1, Canada}
\affiliation{Stewart Blusson Quantum Matter Institute, University of British Columbia, 2355 East Mall, Vancouver, BC V6T~1Z4, Canada}

\author{Dimitrios Koumoulis}
\altaffiliation{Current address: School of Physical Science and Technology, ShanghaiTech University, Pudong, Shanghai 201210, China}
\affiliation{Department of Chemistry and Biochemistry, University of California, Los Angeles, 607 Charles E. Young Drive East, Los Angeles, CA 90095, USA}

\author{C. D. Philip Levy}
\affiliation{TRIUMF, 4004 Wesbrook Mall, Vancouver, BC V6T~2A3, Canada}

\author{Chilin Li}
\affiliation{Shanghai Institute of Ceramics, Chinese Academy of Sciences, 1295 Dingxi Road, Shanghai, P.R. China 200050}

\author{Iain McKenzie}
\affiliation{TRIUMF, 4004 Wesbrook Mall, Vancouver, BC V6T~2A3, Canada}
\affiliation{Department of Chemistry, Simon Fraser University, 8888 University Drive, Burnaby, BC V5A~1S6, Canada}

\author{Rotraut Merkle}
\affiliation{Max-Planck-Institut f{\"u}r Festk{\"o}rperforschung, Heisenbergstra{\ss}e 1, 70569 Stuttgart, Germany}

\author{Gerald D. Morris}
\affiliation{TRIUMF, 4004 Wesbrook Mall, Vancouver, BC V6T~2A3, Canada}

\author{Matthew R. Pearson}
\affiliation{TRIUMF, 4004 Wesbrook Mall, Vancouver, BC V6T~2A3, Canada}

\author{Zaher Salman}
\affiliation{Laboratory for Muon Spin Spectroscopy, Paul Scherrer Institute, CH-5232 Villigen PSI, Switzerland}

\author{Dominik Samuelis}
\altaffiliation{Current address: Heraeus Deutschland GmbH \& Co. KG, Heraeusstra{\ss}e 12--14, 63450 Hanau, Germany}
\affiliation{Max-Planck-Institut f{\"u}r Festk{\"o}rperforschung, Heisenbergstra{\ss}e 1, 70569 Stuttgart, Germany}

\author{Monika Stachura}
\affiliation{TRIUMF, 4004 Wesbrook Mall, Vancouver, BC V6T~2A3, Canada}

\author{Jiyu Xiao}
\affiliation{Department of Chemistry, University of British Columbia, 2036 Main Mall, Vancouver, BC V6T~1Z1, Canada}

\author{Joachim Maier}
\affiliation{Max-Planck-Institut f{\"u}r Festk{\"o}rperforschung, Heisenbergstra{\ss}e 1, 70569 Stuttgart, Germany}

\author{Robert F. Kiefl}
\affiliation{Stewart Blusson Quantum Matter Institute, University of British Columbia, 2355 East Mall, Vancouver, BC V6T~1Z4, Canada}
\affiliation{Department of Physics and Astronomy, University of British Columbia, 6224 Agricultural Road, Vancouver, BC V6T~1Z1, Canada}
\affiliation{TRIUMF, 4004 Wesbrook Mall, Vancouver, BC V6T~2A3, Canada}

\author{W. Andrew MacFarlane}
\email[]{wam@chem.ubc.ca}
\affiliation{Department of Chemistry, University of British Columbia, 2036 Main Mall, Vancouver, BC V6T~1Z1, Canada}
\affiliation{Stewart Blusson Quantum Matter Institute, University of British Columbia, 2355 East Mall, Vancouver, BC V6T~1Z4, Canada}
\affiliation{TRIUMF, 4004 Wesbrook Mall, Vancouver, BC V6T~2A3, Canada}

\date{\today}

\begin{abstract}
We report measurements of the dynamics of isolated \elip\ in single crystal \rutile\ using $\beta$-detected NMR.
From spin-lattice relaxation and motional narrowing, we find two sets of thermally activated dynamics: one below \SI{100}{\kelvin};
and one at higher temperatures.
At low temperature, the activation barrier is \SI{26.8 \pm 0.6}{\milli\electronvolt} with prefactor \SI{1.23 \pm 0.05 e10}{\per\second}.
We suggest this is unrelated to \lip\ motion, and rather is a consequence of electron polarons in the vicinity of the implanted \elip\ that are known to become mobile in this temperature range.
Above \SI{100}{\kelvin}, \lip\ undergoes long-range diffusion as an isolated uncomplexed cation,
characterized by an activation energy and prefactor of \SI{0.32 \pm 0.02}{\electronvolt} and \SI{1.0\pm 0.5 e16}{\per\second},
in agreement with macroscopic diffusion measurements.
These results in the dilute limit from a microscopic probe indicate that \lip\ concentration does not limit the diffusivity even up to high concentrations, but that some key ingredient is missing in the calculations of the migration barrier.
The anomalous prefactors provide further insight into both \lip\ and polaron motion.
\end{abstract}

\maketitle

\section{Introduction \label{sec:introduction}}

\begin{figure}[]
\centering
\includegraphics[width=0.45\textwidth]{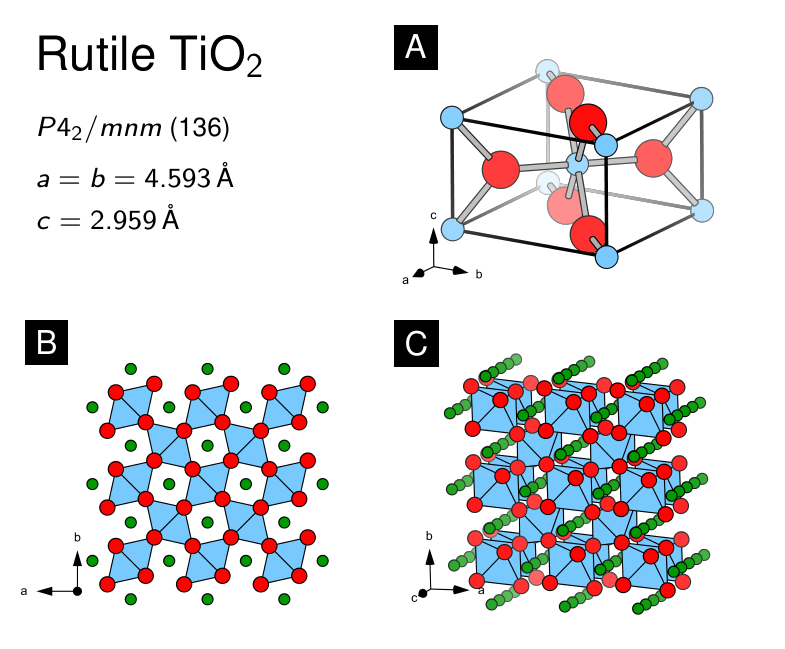}
\caption{\label{fig:rutile-structure}
The rutile structure~\cite{1987-Burdett-JACS-109-3639}.
(A) Titanium-centred conventional unit cell showing Ti (blue) and oxygen (red) ions.
Bonds (grey) are drawn to emphasize coordination.
(B) View along the 4-fold $c$-axis, revealing the channels for fast interstitial \lip\ diffusion. Lithium (green) are shown in the channel centre ($4c$ sites) surrounded by the blue Ti-centred octahedra.
(C) Off-axis view revealing the 1D channels.
}
\end{figure}

The mobility of lithium ions inserted into \rutile\ is exceptionally high and unmatched by any other interstitial cation~\cite{2010-VanOrman-RMG-72-757}.
Even at \SI{300}{\kelvin}, the \lip\ diffusion coefficient is as large as \SI{e-6}{\centi\metre\squared\per\second}~\cite{1964-Johnson-PR-136-A284}, exceeding many state-of-the-art solid-state lithium electrolytes~\cite{2009-Knauth-SSI-180-911,2016-Bachman-CR-116-140}.
Moreover, this mobility is extremely anisotropic~\cite{1964-Johnson-PR-136-A284}, and rutile is a nearly ideal 1D lithium-ion conductor.
This is a consequence of rutile's tetragonal structure~\cite{1987-Burdett-JACS-109-3639}, which has open channels along the $c$-axis that provide a pathway for fast interstitial diffusion (see Fig.~\ref{fig:rutile-structure}).
This has, in part, led to a keen interest in using rutile as an electrode in lithium-ion batteries~\cite{2013-Reddy-CR-113-5364}, especially since the advantages of nanosized crystallites were realized~\cite{2006-Hu-AM-18-1521}.
Simultaneously, much effort has focused on understanding the lithium-ion dynamics~\cite{2001-Koudriachova-PRL-86-1275, 2002-Koudriachova-PRB-65-235423, 2003-Koudriachova-SSI-157-35, 2006-Gligor-SSI-177-2741, 2009-Kerisit-JPCC-113-20998, 2010-Sushko-ECST-28-299, 2012-Yildirim-PCCP-14-4565, 2014-Kerisit-JPCC-118-24231, 2014-Jung-AIPA-4-017104, 2015-Baek-CC-51-15019, 2015-Arrouvel-CTC-1072-43};
however, many underlying details in these studies are inconsistent with available experimental data.
For example, a small activation energy of around \SI{50}{\milli\electronvolt} is consistently predicted~\cite{2001-Koudriachova-PRL-86-1275, 2002-Koudriachova-PRB-65-235423, 2003-Koudriachova-SSI-157-35, 2009-Kerisit-JPCC-113-20998, 2012-Yildirim-PCCP-14-4565, 2014-Kerisit-JPCC-118-24231, 2014-Jung-AIPA-4-017104, 2015-Baek-CC-51-15019, 2015-Arrouvel-CTC-1072-43}, but measured barriers are greater by an order of magnitude~\cite{1964-Johnson-PR-136-A284,2010-Bach-EA-55-4952,2010-Heine-DF-12-95}.
This disagreement is troubling considering the simplicity of both the rutile lattice and the associated \lip\ motion.
One explanation for the discrepancy is that most experimental methods sense the macroscopic ion transport, while theory focuses on elementary microscopic motion.
The two would only be related if the macroscopic transport were not strongly influenced by crystal defects, as might be expected due to the highly one-dimensional mobility.
Thus, for a direct comparison with theory it is important to have microscopic measurements of the \lip\ dynamics.

The electronic properties of rutile are also of substantial interest~\cite{2014-Setvin-PRL-113-086402, 2014-Szot-SSP-65-353}.
While it is natively a wide band-gap (\SI{3}{\electronvolt}) insulator, it can be made an $n$-type semiconductor by introducing electrons into vacant titanium $3d$ $t_{2g}$ orbitals, reducing its valence from $4+$ to $3+$.
This is easily achieved through optical excitation, extrinsic doping, or by oxygen substoichiometry.
Rather than occupying delocalized band states, these electrons form small polarons, where the Ti$^{3+}$ ion is coupled to a substantial distortion of the surrounding oxygen octahedron.
Polaron formation in rutile is not predicted by naive density functional calculations, and to obtain it one must introduce electron interactions~\cite{2007-Deskins-PRB-75-195212}.
Recently, the polaron has been studied optically~\cite{2014-Sezen-SR-4-3808} and by electron paramagnetic resonance (\epr)~\cite{2013-Yang-PRB-87-125201}.
Compared to delocalized band electrons, polaron mobility is quite limited and often exhibits thermally activated hopping.
Calculations predict that the polaron mobility, like interstitial \lip, is also highly anisotropic, with fast transport along the $c$-axis stacks of edge sharing TiO$_{6}$ octahedra~\cite{2007-Deskins-PRB-75-195212, 2009-Kerisit-JPCC-113-20998, 2013-Janotti-PSSRRL-7-199, 2015-Yan-PCCP-17-29949}.
Importantly for our results, the positive charge of an interstitial cation like \lip\ can bind the polaron into a \lip-polaron complex, effectively coupling the electronic and ionic transport~\cite{2012-Shin-SSI-225-590}.
Even before this complex was observed by \epr\ and electron nuclear double resonance (\Endor)~\cite{2013-Brant-JAP-113-053712}, its effect on the mobility of \lip\ was considered theoretically~\cite{2009-Kerisit-JPCC-113-20998, 2012-Yu-JPCL-3-2076}.

To study lithium-ion dynamics in rutile, a technique sensitive to the local environment of \lip\ is desirable.
Nuclear magnetic resonance (\nmr) is a sensitive microscopic probe of matter with a well-developed toolkit for studying ionic mobility in solids~\cite{1948-Bloembergen-PR-73-679, 1979-Richards-TCP-15-141, 1982-Kanert-PR-91-183, 1992-Brinkmann-PNMRS-24-527, 2005-Heitjans-DCM-9-367, 2007-Bohmer-PNMRS-50-87, 2012-Kuhn-SSNMR-42-2, 2012-Wilkening-CPC-13-53, 2016-VinodChandran-ARNMRS-89-1}.
In particular, spin-lattice relaxation (\slr) measurements provide a means of studying fast dynamics.
They are sensitive to the temporal fluctuation in the local fields sensed by \nmr\ nuclei, which induce transitions between magnetic sublevels and relax the ensemble of spins towards thermal equilibrium.
When these stochastic fluctuations induced by, for example, ionic diffusion, have a Fourier component at the Larmor frequency
(typically on the order of \si{\mega\hertz})
\begin{equation} \label{eq:nmr-frequency}
   \omega_{L} = 2 \pi \nu_{L} = 2 \pi \gamma B_{0},
\end{equation}
where $\gamma$ is the gyromagnetic ratio of the \nmr\ nucleus and $B_{0}$ is the applied magnetic field,
the \slr\ rate $\lambda=1/T_{1}$ is maximized.
Complementary information can be obtained from motion induced changes to the resonance lineshape.
In the low temperature limit, the static \nmr\ lineshape is characteristic of the lattice site, with features such as the quadrupolar splitting and
magnetic dipolar broadening from the nuclei of neighboring atoms.
As temperature increases and the hop rate exceeds the characteristic frequency of these spectral features, dynamic averaging
yields substantially narrowed spectra with sharper structure.
This phenomenon is collectively known as ``motional narrowing'' and is sensitive to slow motion with rates typically on the order of \si{\kilo\hertz}.
Together, \slr\ and resonance methods can provide direct access to atomic hop rates over a dynamic range up to nearly \num{6} decades.

Here, we use $\beta$-detected \nmr\ (\bnmr)~\cite{1983-Ackermann-TCP-31-291, 2015-MacFarlane-SSNMR-68-1} to measure the \lip\ dynamics in rutile.
Short-lived \elip\ ions are implanted at low-energies (\SI{\sim 20}{\kilo\electronvolt}) into single crystals of rutile, and their \nmr\ signals are obtained by monitoring the \eli\ nuclear spin-polarization through the anisotropic $\beta$-decay.
$1/T_1$ measurements reveal two sets of thermally activated dynamics: one low-temperature process below \SI{100}{\kelvin} and another at higher temperatures.
The dynamics at high temperature is due to long-range \lip\ diffusion, in agreement with macroscopic diffusion measurements, and corroborated by motional narrowing of the resonance lineshape.
We find a dilute-limit activation barrier of \SI{0.32 \pm 0.02}{\electronvolt}, which is consistent with macroscopic diffusivity, but inconsistent with theory.
We suggest that the dynamics below \SI{100}{\kelvin} and its much smaller activation barrier are related to the low-temperature kinetics of dilute electron polarons.

The paper is organized as follows: Section~\ref{sec:experiment} details the methods used.
The results of spin lattice relaxation and resonance measurements and their analysis form section~\ref{sec:slr} and \ref{sec:resonance}, respectively.
A detailed discussion including comparison to the extensive literature is presented in section~\ref{sec:discussion}, and a summary is given in section~\ref{sec:conclusion}.
The appendices give some further detail on the spin lattice relaxation model (Appendix~\ref{app:rlx}) and on the candidate site for interstitial \lip\ (Appendix \ref{app:site}).

\section{Experiment \label{sec:experiment}}

\bnmr\ experiments were performed at TRIUMF in Vancouver, Canada.
A low-energy \SI{\sim 20}{\kilo\electronvolt} hyperpolarized beam of \elip\ was implanted into rutile single crystals mounted in one of two dedicated spectrometers~\cite{2004-Morris-PRL-93-157601, 2004-Salman-PRB-70-104404, 2015-MacFarlane-SSNMR-68-1, 2014-Morris-HI-225-173}.
The incident ion beam has a typical flux of \SI{\sim e6}{ions\per\second} over a beam spot \SI{\sim 3}{\milli\metre} in diameter.
At these implantation energies, the \elip\ stop at average depths of at least \SI{100}{\nano\meter}~\cite{2014-McFadden-JPCS-551-012032}, as calculated by the \srim\ Monte Carlo code~\cite{2008-Ziegler-SRIM-7}.
Spin-polarization was achieved in-flight by collinear optical pumping with circularly polarized light, yielding a polarization of \SI{\sim 70}{\percent}~\cite{2014-MacFarlane-JPCS-551-012059}.
The samples are one-side epitaxially polished (roughness $<\SI{0.5}{\nano\metre}$), commercial substrates with typical dimensions \SI[product-units=power]{8 x 10 x 0.5}{\milli\metre} (Crystal GmbH, Berlin).
All the samples were transparent to visible light, but straw-coloured, qualitatively indicating a minor oxygen deficiency~\cite{1961-Straumanis-AC-14-493}.
The probe nucleus, \eli, has nuclear spin $I=2$, gyromagnetic ratio $\gamma = \SI{6.3016}{\mega\hertz\per\tesla}$, nuclear electric quadrupole moment $Q=\SI[retain-explicit-plus]{+32.6}{\milli b}$, and radioactive lifetime $\tau_{\beta}=\SI{1.21}{\second}$.

In all the \bnmr\ measurements, the nuclear spin-polarization of \eli\ is monitored through its anisotropic $\beta$-decay, and the observed \emph{asymmetry} of the $\beta$-emissions is proportional to the average longitudinal nuclear spin-polarization~\cite{1983-Ackermann-TCP-31-291, 2015-MacFarlane-SSNMR-68-1}.
The proportionality factor, $A_{0}$, is determined by the $\beta$-decay properties of \eli\ and the detection geometry of each experiment.

\begin{figure}[]
\centering
\includegraphics[width=0.45\textwidth]{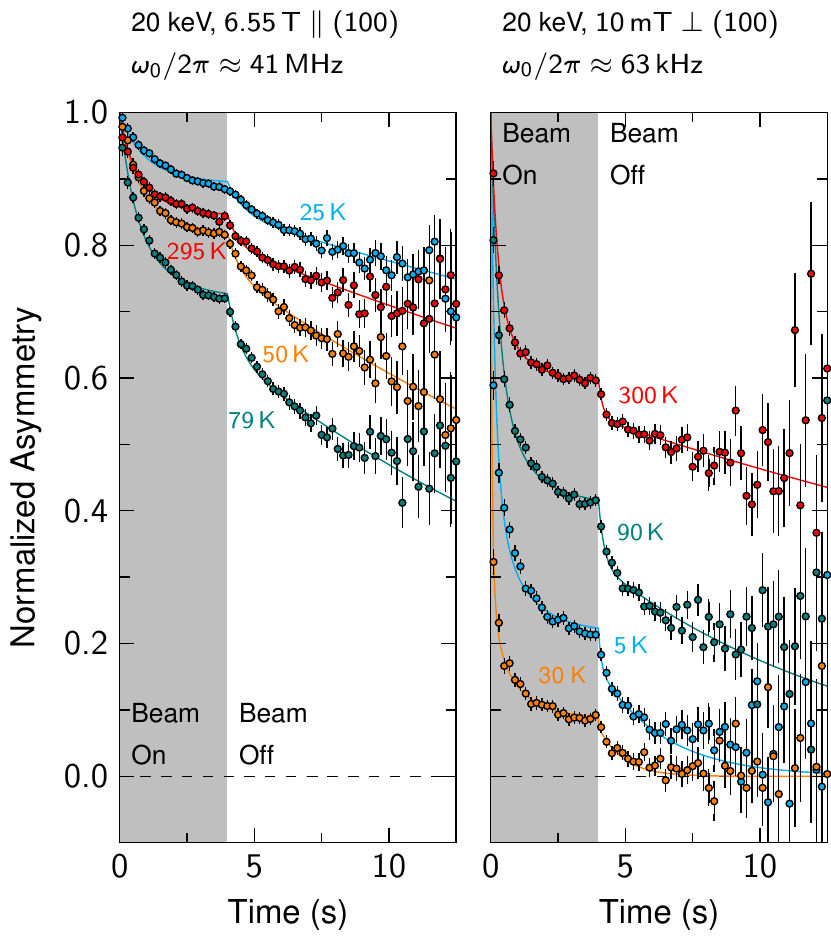}
\caption{\label{fig:slr}
\slr\ spectra of \elip\ in rutile with $B_0 = \SI{6.55}{\tesla}\parallel(100)$ [left] and $\SI{10}{\milli\tesla}\perp(100)$ [right].
The relaxation becomes faster with decreasing magnetic field and is a non-monotonic function of temperature.
The solid lines are a fit to biexponential relaxation function [Eq.~\eqref{eq:biexponential}] convoluted with a \SI{4}{\second} square beam pulse~\cite{2006-Salman-PRL-96-147601,2015-MacFarlane-PRB-92-064409}.
These fit curves were obtained from a global fitting procedure where a temperature independent overall amplitude $A_{0}$ is shared for all spectra at each field, as described in the text.
The spectra have been binned by a factor of \num{20} for clarity.
}
\end{figure}

\slr\ measurements were performed by monitoring the transient decay of spin-polarization both during and following a short pulse of the \elip\ beam~\cite{2006-Salman-PRL-96-147601, 2015-MacFarlane-PRB-92-064409}.
During the pulse, the polarization approaches a steady-state value, while following the pulse, it relaxes to \num{\sim 0}.
This produces a pronounced kink at the end of the beam pulse, characteristic of \bnmr\ \slr\ spectra (see Fig.~\ref{fig:slr}).
Note that no radio-frequency (\rf) field is required for the \slr\ measurements, as the probe spins are implanted in a spin state already far from equilibrium.
As a result, just opposite to conventional \nmr, it is faster and easier to measure \slr\ than to measure the resonance, but as a corollary, this type of relaxation measurement has no spectral resolution.
It represents the spin relaxation of \emph{all} the \eli\ in the sample, not only those corresponding to measurable resonances.
\slr\ rates in rutile were measured from \SIrange[range-phrase=--,range-units=single]{5}{300}{\kelvin} under applied magnetic fields of: \SI{1.90}{\tesla} and \SI{6.55}{\tesla} parallel to the TiO$_{2}$ (100); and \SI{10}{\milli\tesla} and \SI{20}{\milli\tesla} perpendicular to the TiO$_{2}$ (100).

In a continuous \elip\ beam, resonances were acquired in the high-field spectrometer~\cite{2014-Morris-HI-225-173} with a continuous wave (\cw) transverse \rf\ magnetic field stepped slowly through the \eli\ Larmor frequency.
The spin of any on-resonance \eli\ is precessed rapidly by the \rf\ field, resulting in a loss in time-averaged asymmetry.
The evolution of the resonance was recorded over a temperature range of \SIrange{10}{315}{\kelvin} with a highly-homogeneous magnetic field of \SI{6.55}{\tesla} parallel to the TiO$_{2}$ (100).
The resonance frequency was calibrated against that in single crystal MgO (100) at \SI{300}{\kelvin},\cite{2014-MacFarlane-JPCS-551-012033} with the superconducting solenoid persistent. A typical relaxation measurement takes about \SI{20}{\minute}, while a resonance measurement requires about \SI{1}{\hour}.

\section{Results and Analysis \label{sec:results}}

\subsection{Spin-Lattice Relaxation \label{sec:slr}}

Typical \slr\ data at high and low field are shown in Fig.~\ref{fig:slr} for several temperatures.
In high magnetic fields, the relaxation is remarkably fast compared to other oxide insulators~\cite{2014-MacFarlane-JPCS-551-012033, 2003-MacFarlane-PB-326-209, Karner-TBP}, consistent with an earlier report in an intermediate field of \SI{0.5}{\tesla}~\cite{2001-Ogura-HI-136-195}.
It is also immediately evident that the \slr\ rates are strongly dependent on both temperature and field.
The rate of relaxation increases monotonically as the magnetic field is decreased towards zero, see Fig.~\ref{fig:slr}.
At fixed field, however, the temperature dependence of the relaxation is \emph{nonmonotonic}, and there is at least one temperature where the relaxation rate is locally maximized.
Moreover, the temperature of the relaxation rate peak is field-dependent, increasing monotonically with increasing field.

To make these observations quantitative, we now consider a detailed analysis.
The relaxation is not single exponential at \emph{any} field or temperature, but a phenomenological biexponential relaxation function, composed of fast and slow relaxing components yields a good fit.
For an \elip\ ion implanted at time $t'$, the spin polarization at time $t>t'$ follows~\cite{2006-Salman-PRL-96-147601,2015-MacFarlane-PRB-92-064409}:
\begin{equation} \label{eq:biexponential}
   R(t,t^{\prime}) = f_{\text{slow}} e^{- \lambda_{\text{slow}} (t-t^{\prime})} + (1-f_{\text{slow}}) e^{- \lambda_{\text{fast}} (t-t^{\prime})},
\end{equation}
where the rates are $\lambda_{\mathrm{fast}/\mathrm{slow}} = 1/T_{1}^{\mathrm{fast}/\mathrm{slow}}$, and $f_{\text{slow}} \in [0,1]$ is the slow relaxing fraction.
We discuss possible origins for biexponential $R(t,t^{\prime})$ in Appendix~\ref{app:rlx}.

With this model, all the data at each field are fit simultaneously with a shared common initial asymmetry ($A_{0}$) using the \minuit~\cite{1975-James-CPC-10-343} minimization routines within \ROOT~\cite{1997-Brun-NIMA-389-81} to find the optimum global nonlinear least-squares fit.
Notice that the statistical error bars are highly \emph{inhomogenous} with time, characteristic of the radioactive \eli\ decay.
During the beam pulse, the uncertainty decreases with time as the statistics increase, reaching a minimum at the trailing edge of the \SI{4}{\second} beam pulse.
Following the pulse, the error bars grow exponentially as $\exp(-t/\tau_{\beta})$, e.g. Fig.~\ref{fig:slr}.
Accounting for this purely statistical feature of the data is crucial in the analysis.
A subset of the results are shown as solid coloured lines in Fig.~\ref{fig:slr}.
The fit quality is good in each case (global $\tilde{\chi}^{2}\approx 1.1$).
The relaxation rates of the two components are very different with $\lambda_{\mathrm{fast}} > 20 \lambda_{\mathrm{slow}}$, and the analysis distinguishes them clearly.

\begin{figure}[]
\centering
\includegraphics[width=0.45\textwidth]{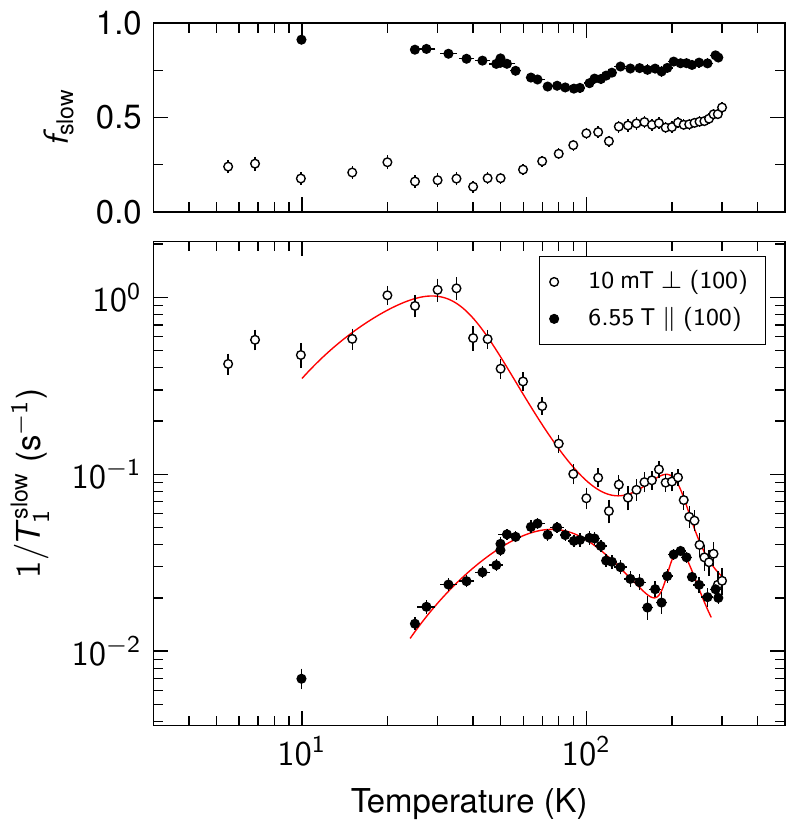}
\caption{\label{fig:slr-slow-fits}
Results from the analysis of the \slr\ measurements using Eq.~\eqref{eq:biexponential} at high- and low-field.
Shown are the fraction of the slow relaxing component (top) and the slow relaxation rate (bottom).
$f_{\text{slow}}$ is surprisingly both temperature- and field-dependent, but increases towards \num{1} by \SI{300}{\kelvin}.
The qualitative features of the \slr\ spectra in Fig.~\ref{fig:slr} can be seen clearly in the two field-dependent rate maxima in $1/T_{1}^{\text{slow}}$.
The solid red line is drawn to guide the eye.
}
\end{figure}

The main fit results are shown in Fig.~\ref{fig:slr-slow-fits}. 
Consistent with the qualitative behaviour of the spectra in Fig.~\ref{fig:slr}, the relaxation rate exhibits two maxima at each field.
This is most apparent in the slow relaxing component at high magnetic field, while at low field, the low temperature peak is substantially broadened.
Generally, a maximum in the relaxation rate occurs when the average fluctuation rate matches the Larmor frequency~\cite{1948-Bloembergen-PR-73-679, 1979-Richards-TCP-15-141, 1982-Kanert-PR-91-183, 1992-Brinkmann-PNMRS-24-527, 2005-Heitjans-DCM-9-367, 2007-Bohmer-PNMRS-50-87, 2012-Kuhn-SSNMR-42-2, 2012-Wilkening-CPC-13-53, 2016-VinodChandran-ARNMRS-89-1}, while the detailed temperature dependence $\lambda(T)$ depends on the character of the fluctuations.

Though the two relaxing components share similarities in their temperature dependence (Fig.~\ref{fig:slr-slow-fits} and Appendix~\ref{app:rlx}), we emphasize that the \emph{slow} relaxing component is the more reliable.
Even though the sample is much larger than the incident ion beamspot, backscattering can result in a small fraction of the 
\elip\ stopping outside the sample which typically produces a correspondingly small fast relaxing asymmetry~\cite{2015-MacFarlane-SSNMR-68-1}.
At high field, where \elip\ relaxation is generally slow, most materials show such a fast relaxing component easily distinguishable from the features of interest;
however, when quadrupolar relaxation is present, which results in multiexponential relaxation for high-spin nuclei~\cite{1970-Hubbard-JCP-53-985, 1982-Becker-ZNA-37-697, 1985-Korblein-JPFMP-15-561} (see Appendix~\ref{app:rlx}), distinguishing the background contribution from an intrinsic fast component becomes difficult.
At low fields, a background is even harder to isolate as \elip\ relaxation is typically fast under these conditions.
Therefore, even though the slow component is a minority fraction at low field (see Fig.~\ref{fig:slr-slow-fits}), we assert that it is the more reliable.  
The fact that, as discussed in the following sections, we are able to reproduce material properties observed with other techniques is strong confirmation of the appropriateness of this choice.

\subsection{Resonance \label{sec:resonance}}

\begin{figure}[]
\centering
\includegraphics[width=0.45\textwidth]{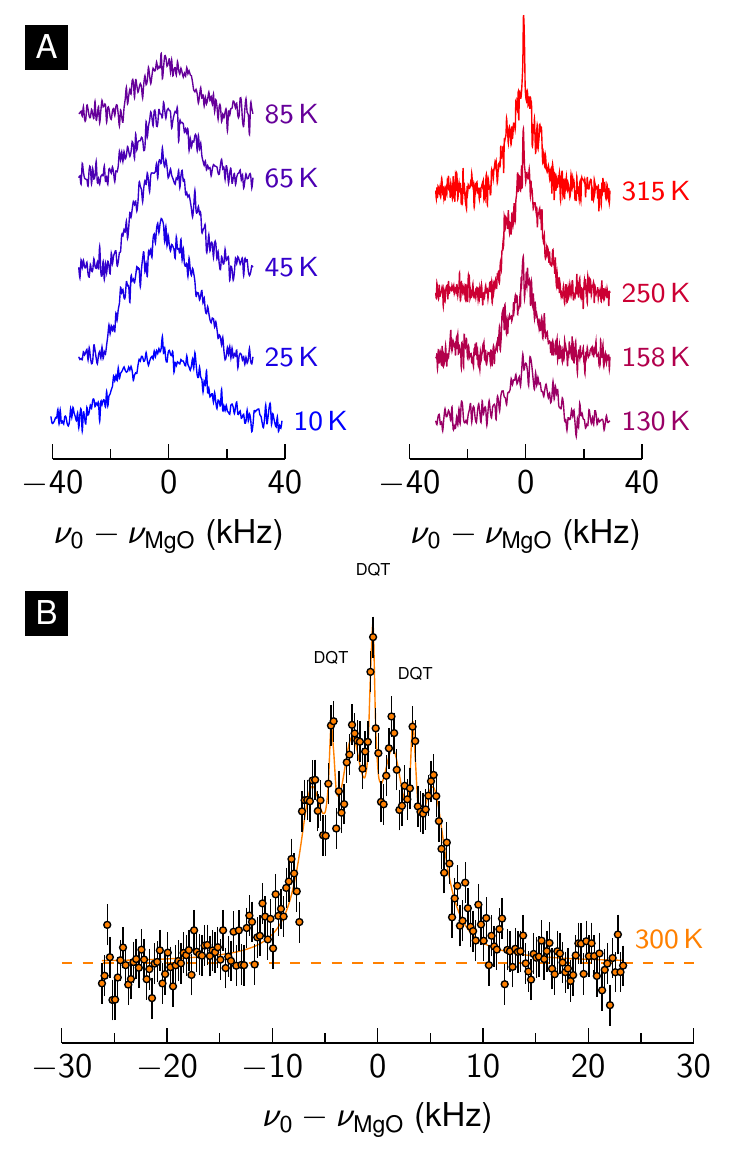}
\caption{\label{fig:resonance-spectra}
\elip\ resonance in \rutile\ with $\SI{6.55}{\tesla}\parallel (100)$.
(A) Temperature dependence of the resonance.
Note that both the vertical and horizontal scales are the same for each spectra, which are offset for clarity.
The zero-shifted position is taken as the resonance frequency of \elip\ in MgO at \SI{300}{\kelvin}.
The lineshape changes substantially with temperature.
Notice that it is most intense at \SI{25}{\kelvin}, but quickly diminishes as the temperature is raised.
At higher temperatures, motional narrowing is apparent and the quadrupolar structure becomes clearly visible.
(B) High-resolution spectrum at \SI{300}{\kelvin}.
The quadrupolar statellite transitions are clearly visible, including narrow double-quantum transitions (DQTs) [positions indicated by DQT] interlaced between the satellites.
The solid orange line is a fit described in the text.
}
\end{figure}

We now turn to the measurements of the \eli\ resonance spectrum at $B_{0}=\SI{6.55}{\tesla}$.
As expected in a noncubic crystal, the \nmr\ is split into a multiplet pattern of quadrupole satellites by the interaction between the \eli\ nucleus and the local electric field gradient (\efg) characteristic of its crystallographic site.
As seen in Fig.~\ref{fig:resonance-spectra}A, the resonance lineshape changes substantially with temperature.
At \SI{10}{\kelvin}, it is broad with an overall linewidth of about \SI{40}{\kilo\hertz}, near the maximum measurable with the limited amplitude \rf\ field of this broadband spectrometer~\cite{2004-Morris-PRL-93-157601, 2014-Morris-HI-225-173}.
Some poorly resolved satellite structure is still evident though.
As the temperature increases, the intensity of the resonance increases considerably with only limited narrowing.
Above \SI{45}{\kelvin}, however, the resonance area decreases dramatically and is minimal near \SI{100}{\kelvin}.
As the temperature is raised further, the quadrupolar splitting becomes more evident, especially above \SI{130}{\kelvin}.
Moreover, the sharpening of these spectral features coincides with a reduction in the breadth of line, along with another increase in signal intensity.
These high-temperature changes are qualitatively consistent with motional narrowing for a mobile species in a crystalline environment.
By room temperature, the spectrum is clearly resolved (see Fig.~\ref{fig:resonance-spectra}B) into the
expected pattern of $2I$ single quantum ($|\Delta m|=1$) quadrupole satellites interlaced with narrower double quantum transitions.
The well-resolved quadrupolar structure indicates a well-defined time-average \efg\ experienced by  a large fraction of the \eli\ at this temperature.
At all temperatures, the centre of mass of the line is shifted to a lower frequency relative to \elip\ in MgO at \SI{300}{\kelvin}.
Note that the resonances in Fig.~\ref{fig:resonance-spectra} are all normalized to the off-resonance steady state asymmetry, which accounts for all the variation of signal intensity due to spin-lattice relaxation~\cite{2009-Hossain-PB-404-914}.

The scale of the quadrupolar interaction is the quadrupole frequency~\cite{1957-Cohen-SSP-5-321}:
\begin{equation} \label{eq:nuq}
\nu_q = \frac{e^2qQ}{4h},
\end{equation}
where the $eq$ is the principal component of the \efg\ tensor.
From the spectra, the splitting is on the order of a few \si{\kilo\hertz}, small relative to the Larmor frequency.
In this limit, the single quantum satellite positions are given accurately by first order perturbation theory as~\cite{1957-Cohen-SSP-5-321}:
\begin{equation} \label{eq:quadrupole}
   \nu_i = \nu_{0} - n_i \frac{\nu_{q}}{2} f(\theta,\phi) ,
\end{equation}
where $n_i=\pm 3 (\pm 1)$ for the outer (inner) satellites, and $f$ is a
function of the polar and azimuthal angles angles $\theta$ and $\phi$ between the external field and \efg\ principal axis system:
\begin{equation} \label{eq:angles}
   f(\theta,\phi) = \frac{1}{2} \left ( 3\cos^{2}\theta -1 + \eta \sin^{2}\theta \cos 2\phi \right). 
\end{equation}
Here, $\eta \in [0,1]$ is the asymmetry parameter for the \efg, which is zero for axial symmetry.
The spectrum thus consists of \num{4} satellites split symmetrically about $\nu_0$.
Unlike the more common case of half-integer spin, there is no unshifted ``main line'' (the $m = \pm 1/2$ transition).
The satellite intensities are also different from conventional \nmr, being determined mainly by the high degree of initial polarization that increases the relative amplitude of the outer satellites~\cite{2014-MacFarlane-JPCS-551-012059}.

\begin{figure}[]
\centering
\includegraphics[width=0.45\textwidth]{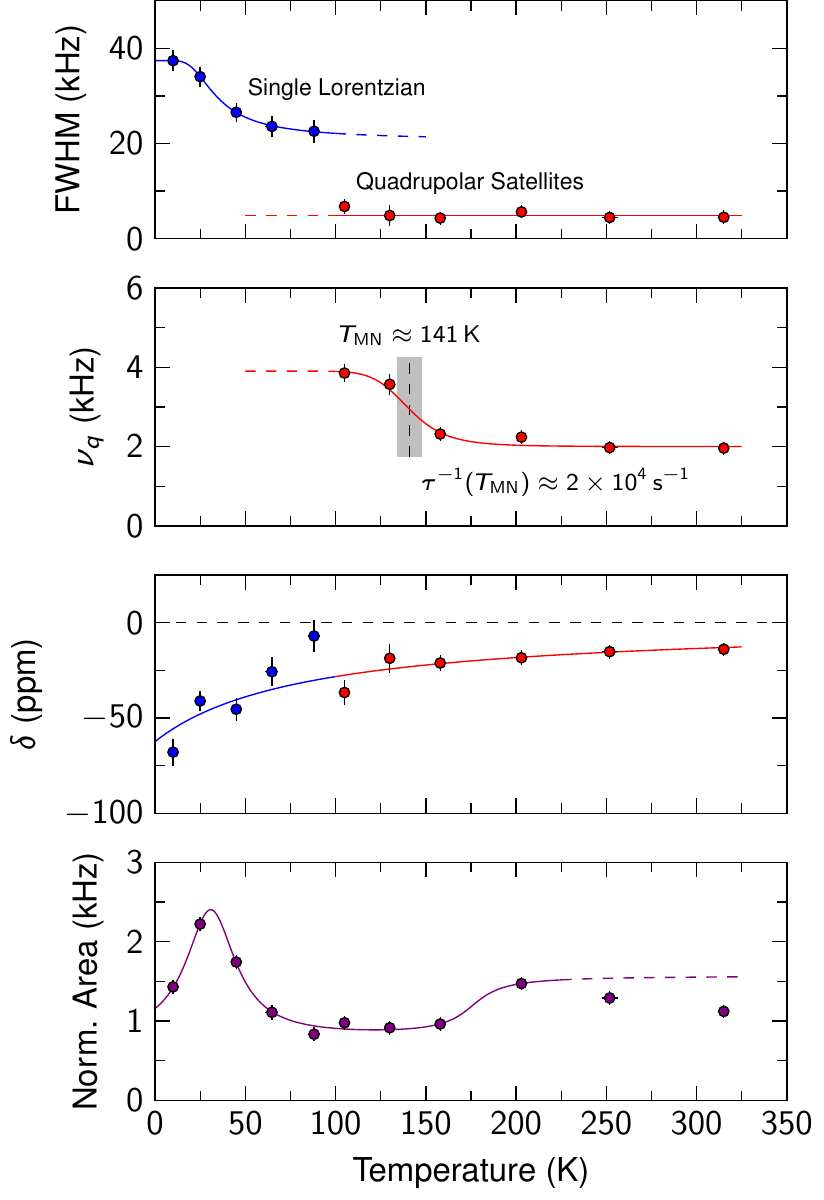}
\caption{\label{fig:resonance-fits}
Results for the analysis of the resonance measurements at high field.
Shown (from top to bottom) are the temperature dependence of the: linewidths, quadrupole frequency, resonance shift, and integrated area. 
In qualitative agreement with the normalized spectra, the resonance is broadest at low temperatures, and gradually reduces in breadth as the temperature is raised.
Above \SI{100}{\kelvin} the quadrupole satellites become resolved and their splitting is gradually reduced by a factor of \num{\sim 2} by \SI{200}{\kelvin}.
An estimate of the motional narrowing temperature $T_{\mathrm{MN}}$ and an associated hop rate $\tau^{-1}(T_{\mathrm{MN}})$ is indicated.
The line is clearly most intense at low temperatures, as indicated by the resonance area, even though it is broadest here.
}
\end{figure}

We now consider a detailed analysis of the resonances.
In agreement with an earlier report~\cite{2014-McFadden-JPCS-551-012032}, the anti-symmetry in helicity-resolved spectra~\cite{2015-MacFarlane-SSNMR-68-1} reveals the resonance is quadrupole split at \emph{all} temperatures;
however, below \SI{100}{\kelvin} the splitting is not well resolved, and an attempt to fit the spectra to a sum of quadrupole satellites proved unsuccessful.
Instead, we use a single Lorentzian in this temperature region to approximate the breadth of the line.
At higher temperatures, where the satellite lines become sharper, a sum of Lorentzians centred at positions given by Eq.~\eqref{eq:quadrupole} (including interlacing double-quantum transitions close to room temperature)~\cite{2015-MacFarlane-SSNMR-68-1} with all $m$-quanta satellites sharing the same linewidth.
From the fits, we extract: the central frequency $\nu_{0}$; the quadrupole splittings $\nu_{q}$; and the overall/satellite linewidths.
From $\nu_{0}$, we calculate the frequency shift $\delta$ relative to \elip\ in MgO at \SI{300}{\kelvin} in parts per million (ppm) using:
\begin{equation} \label{eq:shift}
   \delta = \frac{\nu_{0} - \nu_{\text{MgO}} }{ \nu_{\text{MgO}} }.
\end{equation}
Additionally, the normalized resonance area was estimated following a procedure that removed any effect of \slr\ on the line intensity using a baseline estimation algorithm~\cite{1997-Morhac-NIMA-401-113}.
This allowed for a common integration scheme, independent of a particular fit model.

The results of this analysis are shown in Fig.~\ref{fig:resonance-fits}. 
Though the scatter in the quantities extracted are largest near \SI{100}{\kelvin}, where the resonance is weakest, the qualitative trends noted above are evident.
At low temperatures, the resonance is both widest and most intense, narrowing only modestly approaching \SI{100}{\kelvin}.
The resonance area is clearly largest around \SI{25}{\kelvin}, but quickly diminishes to a minimum around \SI{100}{\kelvin}.
At higher temperatures where the quadrupolar structure is better resolved, the satellites have a nearly temperature independent width of \SI{\sim 4.8}{\kilo\hertz}.
The apparent $\nu_{q}$ reduces gradually from \SI{\sim 3.8}{\kilo\hertz} to nearly half that value by \SI{200}{\kelvin}.
This reduction in splitting coincides with an increase in area,
consistent with the picture of motional averaging of the quadrupolar interaction~\cite{1989-Jansen-JCP-90-1989, 2013-Kuhn-EES-6-3548}.
This implies a fluctuation rate on the order of \SI{\sim 2e4}{\per\second} by \SI{\sim 140}{\kelvin} ($T_{\mathrm{MN}}$ in Fig.~\ref{fig:resonance-fits}).
The resonance shift $\delta$ is both small and negative at all temperatures, gradually increasing towards zero as the temperature is raised.

\section{Discussion \label{sec:discussion}}

The remarkably fast and strongly temperature dependent spin-lattice relaxation at high magnetic field implies an exceptional relaxation mechanism for \eli\ in rutile distinct from other oxide insulators~\cite{2014-MacFarlane-JPCS-551-012033, 2003-MacFarlane-PB-326-209, Karner-TBP}.
The occurrence of a $T_1$ minimum indicates some spontaneous fluctuations are present that are: 1) coupled to the nuclear spin; and 2) their characteristic rate sweeps through the \nmr\ frequency at the temperature of the minimum.
Diffusive motion of \elip\ through the lattice provides at least one potential source of such fluctuations, as is well established in conventional \nmr~\cite{1948-Bloembergen-PR-73-679, 1979-Richards-TCP-15-141, 1982-Kanert-PR-91-183, 1992-Brinkmann-PNMRS-24-527, 2005-Heitjans-DCM-9-367, 2007-Bohmer-PNMRS-50-87, 2012-Kuhn-SSNMR-42-2, 2012-Wilkening-CPC-13-53, 2016-VinodChandran-ARNMRS-89-1};
however, without assuming anything about the particular fluctuations, we extract the temperatures $T_{\mathrm{min}}$ of the two $T_1$ minima (see Fig.~\ref{fig:slr-slow-fits}), by simple parabolic fits, at several magnetic fields corresponding to \nmr\ frequencies spanning three orders of magnitude.
This approach has the advantage of not relying on any particular form of the \nmr\ spectral density function $J(\omega_{L})$~\cite{1988-Beckmann-PR-171-85}, which is proportional to $1/T_{1}$.
Note that there is a clear field dependence to the high-temperature flanks of the $T_{1}$ minima in Fig.~\ref{fig:slr-slow-fits} (i.e., they do no coalesce at high temperatures).
This indicates that the fluctuations are not the result of a 3D isotropic process~\cite{1948-Bloembergen-PR-73-679}, but rather one that is spatially confined to lower dimensions~\cite{1979-Richards-TCP-15-141,1981-Sholl-JPCSSP-14-447}.
Note that for a 1D process (as might be expected in rutile), one needs to account for the characteristic non-Debye fluctuation spectrum~\cite{1977-Fedders-PRB-15-2500, 1978-Fedders-PRB-17-2098}, where only the asymptotic form is known~\cite{1979-Richards-TCP-15-141, 1981-Sholl-JPCSSP-14-447}.

\begin{figure}
\centering
\includegraphics[width=0.45\textwidth]{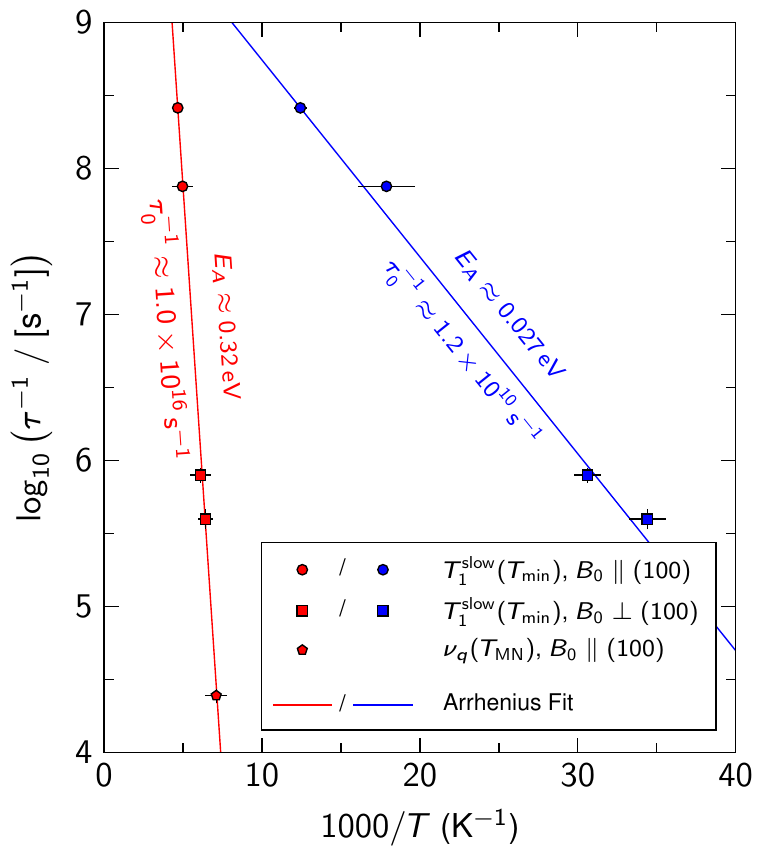}
\caption{\label{fig:arrhenius-tau}
Arrhenius plot of the fluctuation rate extracted from \eli\ \bnmr\ \slr\ and resonance measurements.
Two thermally activated processes can be identified: one below \SI{100}{\kelvin} with a shallow slope (blue points); and one at higher temperatures with a steep slope (red points).
The solid lines are fits to Eq.~\eqref{eq:arrhenius} with the activation energies and prefactors indicated.
}
\end{figure}

Identifying the inverse correlation time of the fluctuations $\tau^{-1}(T_{\mathrm{min}})$ with the \nmr\ frequency $\omega_{L}$ [Eq.~\eqref{eq:nmr-frequency}], we construct an Arrhenius plot of the average fluctuation rate in Fig.~\ref{fig:arrhenius-tau}.
The value of this approach is evident in the linearity of the results which indicates two independent types of fluctuations each with a characteristic activated temperature dependence.
To this plot, we add the estimate of the fluctuation rate causing motional narrowing of the resonance spectra (see $\nu_{q}(T)$ in Fig.~\ref{fig:resonance-fits}), where $\tau^{-1}(T_{\mathrm{MN}})$ matches the static splitting of the line, further expanding the range of $\tau^{-1}$.
That this point lies along the steeper of the two lines is a strong confirmation that the same fluctuations responsible for the high temperature $T_{1}$ minimum cause the motional narrowing.
Linear fits to a simple Arrhenius relationship,
\begin{equation} \label{eq:arrhenius}
   \tau^{-1} = \tau_{0}^{-1} e^{-E_{A}/(k_{B}T)} ,
\end{equation}
where $T$ is the temperature, $k_{B}$ the Boltzmann constant, $E_{A}$ the activation energy, and $\tau_{0}^{-1}$ the prefactor, yield: $\tau_{0}^{-1} = \SI{1.23 \pm 0.05 e10}{\per\second}$ and $E_{A} = \SI{26.8 \pm 0.6}{\milli\electronvolt}$ for the shallow slope, low-temperature fluctuations; and $\tau_{0}^{-1} =\SI{1.0\pm 0.5 e16}{\per\second}$ and $E_{A} = \SI{0.32 \pm 0.02}{\electronvolt}$ for the steep high temperature fluctuations.

The motional narrowing above \SI{100}{\kelvin} is clear evidence that the corresponding fluctuations are due to long-range diffusive motion of \elip.
Unlike liquids, where motion causes the broad solid state lines to collapse to a single narrow Lorentzian, fast interstitial diffusion in a crystal averages only some of the features of the lineshape, e.g., see the $^7$Li spectra in Li$_3$N~\cite{1981-Messer-JPCSSP-14-2731}.
In particular, since the quadrupole splitting (the major spectral feature of the \eli\ resonance in rutile), is finite at every site, fast motion between sites results in an averaged lineshape consisting of quadrupole satellites split by an average \efg\ of reduced magnitude, see Fig.~\ref{fig:resonance-spectra}.
From this we conclude that the rate $\tau^{-1}$ for the steep high temperature fluctuations in Fig.~\ref{fig:arrhenius-tau} should be identified with the rate of activated hopping of \elip\ between adjacent sites --- the elementary atomic process of diffusion in a crystal lattice.
Further confirmation of this identification comes from the excellent agreement of the activation energy with macroscopic diffusion measurements based on optical absorption~\cite{1964-Johnson-PR-136-A284} and impedance spectroscopy~\cite{2010-Bach-EA-55-4952}.

\begin{figure}
\centering
\includegraphics[width=0.45\textwidth]{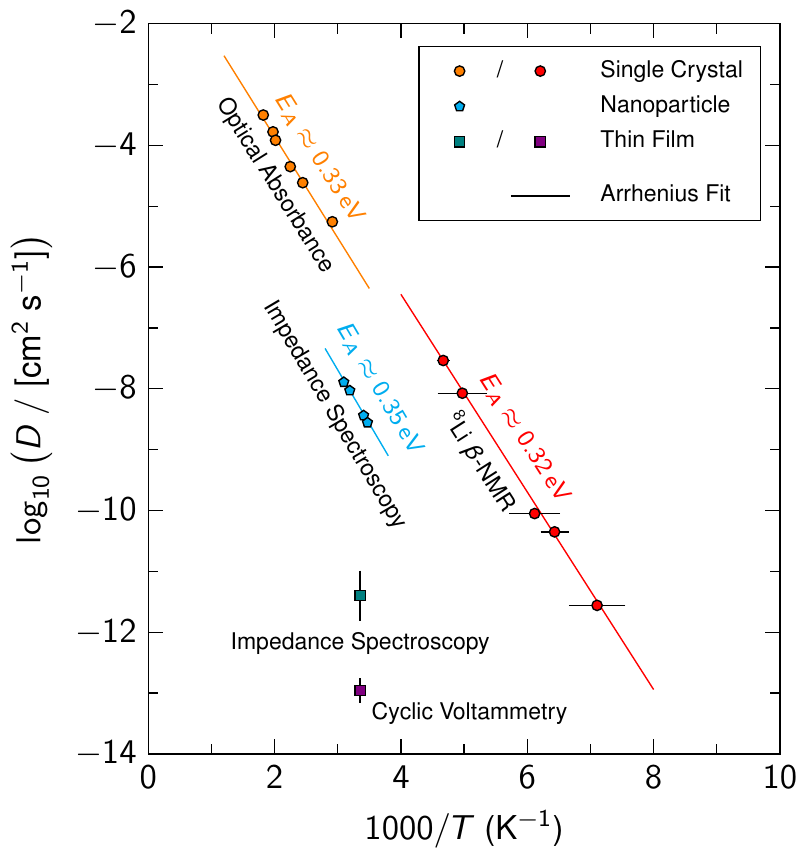}
\caption{\label{fig:arrhenius-diffusion}
Arrhenius plot of the \lip\ diffusion coefficient in rutile estimated from the \elip\ hop rate extracted in Fig.~\ref{fig:arrhenius-tau} using the Einstein-Smolouchouski expression [Eq.~\eqref{eq:einstein-smoluchouski}] with $f=1$, $d=1$, and $l=\SI{1.5}{\angstrom}$.
Literature values obtained from rutile single crystals~\cite{1964-Johnson-PR-136-A284}, nanoparticles~\cite{2010-Bach-EA-55-4952}, and thin films~\cite{2004-Churikov-RJE-40-63,2014-Churikov-JSSE-18-1425} are shown for comparison.
Note that the concentration of lithium varies greatly in the reported values.
While a clear deviation in the magnitude of $D$ is observed for different forms of rutile, the similarity of $E_{A}$ implies a common diffusion mechanism.
}
\end{figure}

For a closer comparison with these experiments, we convert our hop rates to diffusivity via the Einstein-Smolouchouski expression:
\begin{equation} \label{eq:einstein-smoluchouski}
   D = f\frac{l^{2}}{2d\tau},
\end{equation}
where $l$ is the jump distance, $d=1$ is the dimensionality, and $f$ is the correlation factor, assumed to be unity for direct interstitial diffusion.
Using $l\approx\SI{1.5}{\angstrom}$, based on the ideal rutile lattice (details of the precise site of \elip\ are discussed below), and compare the results to $D$ measured by other methods in single crystal~\cite{1964-Johnson-PR-136-A284}, thin film~\cite{2004-Churikov-RJE-40-63,2014-Churikov-JSSE-18-1425}, and nanocrystalline~\cite{2010-Bach-EA-55-4952} rutile in Fig.~\ref{fig:arrhenius-diffusion}.
The agreement in activation energy is apparent in the similarity of the slopes, but our $D$ is somewhat larger than the macroscopic diffusivity due to a larger prefactor.
In our measurements \elip\ is essentially in the dilute limit, while the bulk measurements have much higher concentrations.
One might expect repulsive $\mathrm{Li}^{+}\mathrm{-}\mathrm{Li}^{+}$ interactions would inhibit ionic transport, and yield a smaller macroscopic $D$;
however, the agreement in $E_A$ with the macroscopic $D$, where the (Li/Ti) concentration is as high as \SI{30}{\percent}~\cite{2010-Bach-EA-55-4952}, implies the barrier is very insensitive to concentration, probably due to strong screening of the Coulomb interaction by the high dielectric response of rutile~\cite{1961-Parker-PR-124-1719}.
Alternatively, Eq.~\eqref{eq:einstein-smoluchouski} shows that either overestimating the jump distance $l$ or the presence of correlated hopping (that reduces $f$ from unity)~\cite{2007-Mehrer-DIS-7-105} would lead to an overestimate of $D$ and might account for some of the discrepancy.
Note that isotopic mass effects on $D$ are expected to be negligible~\cite{1966-Johnson-JAP-37-668}.

While our $E_A$ agrees well with macroscopic measurements~\cite{1964-Johnson-PR-136-A284, 2010-Bach-EA-55-4952}, it disagrees with theory by nearly an order of magnitude~\cite{2001-Koudriachova-PRL-86-1275,2002-Koudriachova-PRB-65-235423,2003-Koudriachova-SSI-157-35,2006-Gligor-SSI-177-2741,2009-Kerisit-JPCC-113-20998,2010-Sushko-ECST-28-299,2012-Yildirim-PCCP-14-4565,2014-Kerisit-JPCC-118-24231,2014-Jung-AIPA-4-017104,2015-Baek-CC-51-15019,2015-Arrouvel-CTC-1072-43}.
Generally $E_A$ is a more robustly determined quantity than the absolute value of $D$ at a single temperature, which may exhibit dependence on both sample~\cite{1964-Johnson-PR-136-A284} and measurement technique, e.g., Fig.~\ref{fig:arrhenius-diffusion}.
Moreover, several calculations also predict a strong $E_{A}$ dependence on \lip\ concentration~\cite{2006-Gligor-SSI-177-2741, 2012-Yildirim-PCCP-14-4565, 2014-Jung-AIPA-4-017104}, inconsistent with our results.
While a concentration dependence to $D$ has been observed~\cite{1964-Johnson-PR-136-A284, 2010-Bach-EA-55-4952}, it must find an explanation other than a change in $E_{A}$.
From this we conclude that some ingredient is missing in the theoretical treatments, possibly related to the lattice relaxation around interstitial \lip\ that has a strong effect on the calculated barrier~\cite{2002-Koudriachova-PRB-65-235423,2003-Koudriachova-SSI-157-35}.
Our result is also inconsistent with the suggestion~\cite{2009-Kerisit-JPCC-113-20998} that the higher barrier is characteristic of diffusion of the \lip -polaron complex instead of simply interstitial \lip.
We discuss this point at more length below.

We turn now to the fluctuations that predominate below \SI{100}{\kelvin} and cause the low temperature $T_{1}$ minimum.
While, we cannot be as conclusive about their origin, we delineate some interesting possibilities.
In contrast to the long-range diffusive behaviour at higher temperature, the small activation energy we find is in the range of barriers obtained from molecular dynamics~\cite{2009-Kerisit-JPCC-113-20998,2012-Yildirim-PCCP-14-4565,2014-Kerisit-JPCC-118-24231,2015-Arrouvel-CTC-1072-43} and density functional~\cite{2001-Koudriachova-PRL-86-1275,2002-Koudriachova-PRB-65-235423,2003-Koudriachova-SSI-157-35,2012-Yildirim-PCCP-14-4565,2014-Jung-AIPA-4-017104,2015-Baek-CC-51-15019} calculations for interstitial \lip;
however, this appears to be coincidental, since the absence of motional narrowing in this temperature range is inconsistent with long-range motion.

On the other hand, the relaxation may be caused by some highly localized \lip\ motion at low temperature.
Local dynamics of organic molecules in solids are well-known, for example the rotation of methyl groups of molecules intercalated into crystalline hosts, where they can cause some limited dynamic averaging of the \nmr\ lineshape~\cite{1989-Jansen-JCP-90-1989} and relaxation~\cite{1984-Muller-Warmuth-CS-11-1}.
Analogous effects are found for some point defects in crystals.
For example, a small substitutional cation may adopt one of several equivalent off-centre sites surrounding the high symmetry site of the large missing host cation, and subsequently hop randomly among these sites within the anionic cage, e.g., Ag$^+$ in RbCl~\cite{1982-Kanert-PR-91-183}.

To expand further on this possibility, we now consider the \elip\ site in rutile in more detail.
When Li is introduced either thermally or electrochemically, it is known to occupy the open channels along the $c$-axis.
Two high-symmetry sites are available here: the Wyckoff $4c$ site within a distorted oxygen octahedron; and the $4d$ quasi-tetrahedral site, but the precise location remains controversial and may depend on Li concentration~\cite{2006-Hu-AM-18-1521, 2008-Borhhols-CM-20-2949, 2009-Vijayakumar-JPCC-113-14567}.
Although ion implantation is far from a thermal process, the implanted ion often stops in the most energetically stable site in the unit cell.
From first-principles, the lowest energy site for isolated \lip\ is $4c$ along the centre of the $c$-axis channel~\cite{2009-Kerisit-JPCC-113-20998, 2012-Yildirim-PCCP-14-4565, 2014-Jung-AIPA-4-017104, 2015-Baek-CC-51-15019}, see Fig.~\ref{fig:rutile-structure}B and C.
In disagreement with these calculations, an off-centre site near $4c$ has been predicted~\cite{2001-Koudriachova-PRL-86-1275, 2002-Koudriachova-PRB-65-235423, 2003-Koudriachova-SSI-157-35}, but this seems unlikely given the modest size of the quasi-octahedral cage compared to the \lip\ ionic radius~\cite{1976-Shannon-AC-A32-751}.
Metastable sites outside the channels (in the stacks of TiO$_6$ octahedra) have substantially higher energies~\cite{2001-Koudriachova-PRL-86-1275, 2002-Koudriachova-PRB-65-235423, 2003-Koudriachova-SSI-157-35, 2015-Arrouvel-CTC-1072-43} and would also be characterized by much larger \efg s and quadrupole splittings (in Appendix~\ref{app:site}, we consider the prospects for using the quadrupole splitting to \emph{determine} the \elip\ site).
If, on the other hand, \elip\ stops at a metastable site along the channels, such as $4d$, it would have a very small barrier to moving to the nearest $4c$ site.
Thus, while we cannot rule out some local motion of \elip\ at low temperature, we regard it as unlikely.
Moreover, it is not clear how local motion could account for the temperature evolution of the resonance area, whose main feature is a peak in intensity below \SI{50}{\kelvin}, see Fig.~\ref{fig:resonance-spectra}.

We now consider another source of low $T$ fluctuations, namely the electron polaron, which, for simplicity, we denote as Ti$'_{\mathrm{Ti}}$.
The polaron is only slightly lower in energy (\SI{0.15}{\electronvolt}) than the delocalized electronic state ($e^{\prime}$) at the bottom of the rutile conduction band~\cite{2014-Setvin-PRL-113-086402, 2013-Janotti-PSSRRL-7-199}.
We can write the localization transition as:
\begin{equation*}
   \mathrm{Ti}_{\mathrm{Ti}} + e^{\prime} \rightleftharpoons \mathrm{Ti}_{\mathrm{Ti}}^{\prime}.
\end{equation*}
Having localized, the polaron can migrate in an activated manner, with a calculated $E_A$ that may be as low as \SI{\sim 30}{\milli\electronvolt} for adiabatic hopping along the $c$-axis stacks of TiO$_6$ octahedra~\cite{2007-Deskins-PRB-75-195212, 2009-Kerisit-JPCC-113-20998, 2013-Janotti-PSSRRL-7-199, 2015-Yan-PCCP-17-29949}.
Note that polaron localization also results in the formation of a local electronic magnetic moment --- the polaron is a \emph{paramagnetic defect} --- as is clearly confirmed by \epr~\cite{2013-Yang-PRB-87-125201}.
At low temperature, the polaron is likely weakly bound to other defects such as an oxygen vacancies, from which it is easily freed~\cite{1967-Dominik-PR-163-756, 1996-Yagi-PRB-54-7945}.
If the one dimensionally mobile polaron and interstitial \lip\ on adjacent sublattices come into close proximity, they may form a bound state:
\begin{equation*}
   \mathrm{Li}_{i}^{\cdot} + \mathrm{Ti}_{\mathrm{Ti}}^{\prime}  \rightleftharpoons  \mathrm{Li}_{i}^{\cdot}\mathrm{-}\mathrm{Ti}_{\mathrm{Ti}}^{\prime},
\end{equation*}
that is a charge-neutral paramagnetic defect complex that has been characterized by \epr\ and \Endor~\cite{2013-Brant-JAP-113-053712}.
The complex is predicted to be quite stable~\cite{2009-Kerisit-JPCC-113-20998}, but its \epr\ signal broadens and disappears above about \SI{50}{\kelvin}~\cite{2013-Brant-JAP-113-053712}.
The complex is also expected to be mobile via a tandem hopping process~\cite{2009-Kerisit-JPCC-113-20998}.

\elip\ bound to a polaron will have a very different \nmr\ spectrum than the isolated interstitial.
The $3+$ charge of the nearby $\mathrm{Ti}_{\mathrm{Ti}}^{\prime}$ will alter the \efg\ and modify the quadrupole splitting, but the magnetic hyperfine field of the unpaired electron spin is an even larger perturbation, so strong in fact, that complexed \elip\ will not contribute at all to the resonances in Fig.~\ref{fig:resonance-spectra}, since, based on the \Endor~\cite{2013-Brant-JAP-113-053712} their resonance frequency is shifted by at least \SI{350}{\kilo\hertz}.
For this reason we also exclude the possibility that the high temperature dynamics corresponds to motion of the $\mathrm{Li}_{i}^{\cdot}\mathrm{-}\mathrm{Ti}_{\mathrm{Ti}}^{\prime}$ complex.
There is no evidence that its spin polarization is wiped out by fast relaxation which would result in a missing fraction in Fig.~\ref{fig:slr}.
However, if immediately after implantation the \elip\ is free for a time longer than the period of precession in the \rf\ field (\SI{\sim 1}{\kilo\hertz}), it will contribute to the resonance before binding with a polaron.
Similarly, if the $\mathrm{Li}_{i}^{\cdot}\mathrm{-}\mathrm{Ti}_{\mathrm{Ti}}^{\prime}$ complex undergoes cycles of binding and unbinding at higher temperature, provided it is unbound for intervals comparable to the precession period, it will participate in the resonance.
In analogy with the closely related technique of \rf\ muon spin resonance (\rf-\musr)~\cite{1991-Kreitzman-HI-65-1055}, one can thus use the resonance amplitude of the diamagnetic \elip\ in Fig.~\ref{fig:resonance-spectra} to follow kinetic processes involving the implanted ion, e.g., the hydrogenic muonium defect in silicon~\cite{1995-Kreitzman-PRB-51-13117}.
Along these lines, we suggest that the nonmonotonic changes in resonance amplitude at low temperature reflect dynamics of the \lip-polaron complexation.
The sample in our measurements is nominally undoped, and we expect the main source of polarons is oxygen substoichiometry.
From its color~\cite{1961-Straumanis-AC-14-493}, it could have oxygen vacancies at the level of \SI{0.1}{\percent} or less and polarons resulting from these vacancies are known to become mobile below \SI{50}{\kelvin}~\cite{1967-Dominik-PR-163-756, 1996-Yagi-PRB-54-7945}.
Alternatively, polarons may result from electron-hole excitations created by the implantation of \elip.

The large increase in resonance area between \num{10} and \SI{25}{\kelvin} implies some form of slow dynamics on the timescale of the \eli\ lifetime $\tau_{\beta}$.
This could be a modulation of the \efg, but more likely it is a magnetic modulation related to the polaron moment as it mobilizes.
This is not motional narrowing, but rather a slow variation in the resonance condition, such that the applied \rf\ matches the resonance frequency for many more \eli\ at some point during their lifetime.
The increase of intensity then corresponds to the onset of polaron motion, while the loss in intensity with increasing temperature is due to formation of the \lip-polaron complex, and the fluctuations from this motion also become fast enough to produce the $T_{1}$ minimum.
The complex does not survive to high temperatures, though, and, based on the resonance intensity, the motionally narrowed quadrupolar split resonance at high temperature corresponds to nearly all of the \eli.
The $E_{A}$ from the low temperature slope in Fig.~\ref{fig:arrhenius-tau} is remarkably compatible with the thermal instability of the intrinsic (unbound) polaron in rutile~\cite{2007-Deskins-PRB-75-195212, 2009-Kerisit-JPCC-113-20998, 2013-Janotti-PSSRRL-7-199, 2015-Yan-PCCP-17-29949, 2013-Brant-JAP-113-053712, 2013-Yang-PRB-87-125201}, consistent with this picture.

Aside from the activation energies, the prefactors $\tau_{0}^{-1}$ from Eq.~\eqref{eq:arrhenius} may provide further information on the processes involved.
For atomic diffusion, the prefactor is often consistent with a vibrational frequency of the atom in the potential well characteristic of its crystalline site, typically \SIrange[range-phrase=--,range-units=single]{e12}{e13}{\per\second}.
Prefactor \emph{anomalies} refer to any situation where $\tau_{0}^{-1}$ falls outside this range~\cite{1983-Villa-SSI-9-1421}.
From Fig.~\ref{fig:arrhenius-tau}, we see that $\tau_{0}^{-1}$ for the high temperature dynamics is anomalously high, while for the low temperature process it is anomalously low.
Within thermodynamic rate theory~\cite{1949-Wert-PR-76-1169, 1957-Vineyard-JPCS-3-121}:
\begin{equation}
   \tau_{0}^{-1} = \tilde{\nu}_{0} e^{\Delta S/k_B},
\end{equation}
where $\Delta S$ is the entropy of migration~\cite{1989-Dobson-PRB-40-2962}, and $\tilde{\nu}_{0}$ is the attempt frequency.
For closely related processes, $\Delta S$ is \emph{not} independent of $E_{A}$, giving rise to Meyer-Neldel (enthalpy-entropy) correlations between the Arrhenius slope and intercept~\cite{1987-Almond-SSI-23-27, 1992-Yelon-PRB-46-12244, 2006-Yelon-RPP-69-1145}, but independent of such correlations, a prefactor anomaly may simply result from $\Delta S/k_B$ being substantially different from \num{1}.

We first consider the high temperature prefactor, noting that the bulk diffusivity also shows an unusually large $D_0$~\cite{1964-Johnson-PR-136-A284}.
Prefactors of this magnitude are uncommon, but not unprecedented.
For example, $^7$Li \nmr\ in LiF at high temperature yields a comparably large $\tau_{0}^{-1}$ for vacancy diffusion in the ``intrinsic'' region~\cite{1963-Eisenstadt-PR-132-630}.
Similarly, a large prefactor is observed from $^{19}$F \nmr\ in superionic PbF$_2$~\cite{1977-Boyce-SSC-21-955}.
The latter case was attributed, not to motion of an isolated fluoride anion, but rather to the total effect of all the mobile interstitial F$^{-}$, whose concentration is also activated.
This may also explain the LiF prefactor, but it clearly does not apply to the extrinsic implanted \elip\ in the dilute limit.
With the advent of sensitive atomic resolution probes of surfaces in the past few decades, a very detailed picture of diffusion on crystal surfaces has emerged~\cite{2005-Tringides-DCM-7-285, 2007-Antczak-SSR-62-39}, which can help to refine our ideas about bulk diffusion.
For example, in some cases, the Arrhenius prefactor of adatoms diffusing along a step edge is significantly enhanced over a flat terrace~\cite{1995-Roder-PRL-74-3217}.
Like the channels in rutile, step edges consist of a 1D array of vacant sites, but the direct relevance is not clear, since the adatoms are generally far from the dilute limit.
We suggest that most reasonable explanation of the high $\tau_{0}^{-1}$ for long-range diffusion of \lip\ in rutile is a large $\Delta S$ which can result from a ballistic picture of hopping~\cite{1989-Dobson-PRB-40-2962}.
Similarly, a (Li/Ti) concentration dependence of $\Delta S$ may contribute to the observed concentration dependence of $D$~\cite{1964-Johnson-PR-136-A284, 2010-Bach-EA-55-4952}.

A more common and widely discussed case is a small prefactor as we find at low temperature.
Low prefactors are often encountered in superionic conductors both in \nmr~\cite{1979-Richards-TCP-15-141, 1979-Boyce-PR-51-189} and transport measurements, where they have been attributed to a breakdown of rate theory~\cite{1978-Huberman-SSC-25-759} or to low dimensionality that is often found for these structures~\cite{1978-Richards-SSC-25-1019} (the latter certainly applies to rutile).
However, as argued above, the low temperature \eli\ relaxation likely reflects polaron dynamics rather than \lip\ motion.
Evidence for this comes from the low temperature evolution of the electronic conductivity of lightly deoxidized rutile that shows a resistivity minimum at about \SI{50}{\kelvin}~\cite{1996-Yagi-PRB-54-7945}.
The complex low temperature behavior of rutile probably combines polaron binding to defects~\cite{1967-Dominik-PR-163-756, 1996-Yagi-PRB-54-7945} with intrinsic polaronic conductivity and the instability to delocalize~\cite{2013-Janotti-PSSRRL-7-199}.
Prefactors for defect-bound polarons at low temperature are significantly lower than our $\tau_{0}^{-1}$~\cite{1967-Dominik-PR-163-756}.
It would be interesting to compare our prefactor with the activated disappearance of the \epr~\cite{2013-Brant-JAP-113-053712, 2013-Yang-PRB-87-125201} to test the connection between these two phenomena with very similar activation energies.

\section{Conclusion \label{sec:conclusion}}

In summary, using low-energy ion-implanted \eli\ \bnmr, we have studied the dynamics of isolated \elip\ in \rutile.
Two sets of thermally activated dynamics were found: one below \SI{100}{\kelvin}; and one at higher-temperatures.
At low temperature, an activation barrier of \SI{26.8 \pm 0.6}{\milli\electronvolt} is measured with an associated prefactor of \SI{1.23 \pm 0.05 e10}{\per\second}.
We suggest this is unrelated to \lip\ motion, and rather is a consequence of electron polarons in the vicinity of the implanted \elip\ that are known to become mobile in this temperature range.
Above \SI{100}{\kelvin}, \lip\ (not polaron complexed) undergoes long-range diffusion, characterized by an activation energy and prefactor of \SI{0.32 \pm 0.02}{\electronvolt} and \SI{1.0\pm 0.5 e16}{\per\second}, in agreement with macroscopic measurements.
These results in the dilute limit from a microscopic probe indicate that \lip\ concentration does not limit the diffusivity even up to high concentrations, but that some key ingredient is missing in the calculations of the barrier.
Low temperature polaronic effects may also play a role in other titanate \lip\ conductors, such as the perovskites~\cite{1999-Emery-JPCM-11-10401} and spinels~\cite{2017-Sugiyama-PRB-96-094402}. 
The present data, combined with \epr\ and transport studies, will further elucidate their properties.

\appendix

\section{Biexponential Relaxation \label{app:rlx}}

\begin{figure}
\centering
\includegraphics[width=0.45\textwidth]{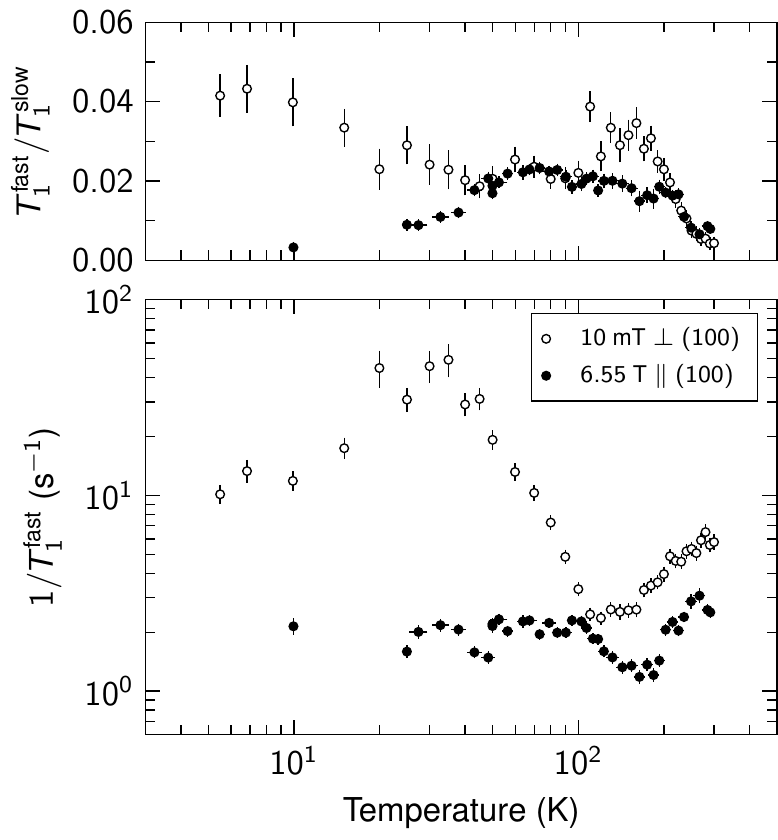}
\caption{\label{fig:slr-fast-fits}
Results from the analysis described in Section~\ref{sec:slr} of the \slr\ measurements using Eq.~\ref{eq:biexponential} at high- and low-field.
Shown are the ratio of the fast/slow relaxation times (top) and the fast relaxation rate (bottom).
The ratio varies weakly with temperature and is field independent for much of it, deviating only in the vicinity of $T_{1}^{\mathrm{slow}}(T_{\mathrm{min}})$.
Some of the features clearly seen in $1/T_{1}^{\mathrm{slow}}$ are apparent in $1/T_{1}^{\mathrm{fast}}$ (cf. Fig.~\ref{fig:slr}), but much less pronounced, especially at high field.
}
\end{figure}

Following the analysis described in Section~\ref{sec:slr}, the fast relaxing component extracted from fitting the \slr\ measurements at high- and low-field to a biexponential relaxation [Eq.~\eqref{eq:biexponential}] is shown in Fig.~\ref{fig:slr-fast-fits}.
While some of the qualitative features seen in the slow component at low temperatures are apparent in $1/T_{1}^{\mathrm{fast}}$, they are much less pronounced at high field (see Fig.~\ref{fig:slr-slow-fits}).
The monotonic increase in $1/T_{1}^{\mathrm{fast}}$ above \SI{\sim 150}{\kelvin} contrasts the behaviour of the slow component  and dominates over any local maxima that may be present.
Interestingly, the ratio of relaxation times $T_{1}^{\mathrm{fast}}/T_{1}^{\mathrm{slow}}$ varies only weakly with temperature and remains field-independent over much of the temperature range.
Deviations from a field-independent ratio occur near the local maxmina in $1/T_{1}^{\mathrm{slow}}$, clearly visible in Fig.~\ref{fig:slr-slow-fits}.

We now consider what might produce the biexponential relaxation $R(t,t')$ [Eq.~\eqref{eq:biexponential}].
A fraction of the implanted \elip\ stopping at a metastable crystallographic site at low temperature would show distinct resonance and relaxation, e.g., \elip\ in simple metals~\cite{2009-Hossain-PB-404-914}.
This would, however, be independent of applied field and would exhibit a very different temperature dependence from the in-channel diffusing site.
As we find no clear evidence for multiple sites in the resonance analysis or field-independent activated modulation of the \slr\ rates, we conclude a secondary site cannot be the source of the biexponential relaxation.

If quadrupolar fluctuations are the dominant source of relaxation, as would be expected for \elip\ diffusion, then on the low temperature side of the $T_{1}$ minimum, where the fluctuations are slow compared to $\omega_{L}$, the relaxation may be intrinsically biexponential for spin $I=2$~\cite{1982-Becker-ZNA-37-697, 1985-Korblein-JPFMP-15-561}.
However, these Redfield-theory calculations of $R(t,t')$ differ in several key assumptions from our situation, specifically: 1) the initial state of the optically polarized \eli\ spin is quite different~\cite{2014-MacFarlane-JPCS-551-012059}; 2) we are not always in the extreme high field limit; and 3) one dimensional hopping yields a non-Debye fluctuation spectrum~\cite{1977-Fedders-PRB-15-2500, 1978-Fedders-PRB-17-2098}.
Moreover, above the $T_{1}$ minimum, where the fluctuations are fast, the biexponential should collapse to a single exponential~\cite{1982-Becker-ZNA-37-697}, which is certainly not the case here, particularly for the low field data.
It remains to be seen whether a suitably modified theory along these lines could account for the field dependence of the biexponential at high temperatures.

In contrast, at low temperature, we suspect the relaxation has a significant, possibly dominant, contribution from the polaron magnetic moment.
Here, the spectrum of magnetic field fluctuations is naturally field dependent, yielding a strongly field dependent $\lambda$.
In this case, some of the fast component at low fields would cross over to the high field slow component as $\lambda(B_0)$ decreases with increasing field.
A detailed field dependence of the relaxation at low field would help to confirm this.
The character of the relaxation (magnetic, quadrupolar or mixed) can also be tested by comparison with another Li isotope, such as the spin $3/2$ $^9$Li as has been recently demonstrated~\cite{2017-Chatzichristos-PRB-96-014307}.
It is important to recall that the amplitude of the relaxing signal is temperature independent even at low field, meaning that we do not have a fraction of the signal that is so fast relaxing that it is lost (wiped out) as has been seen in $^7$Li \nmr\ from polarons in the perovskite Li$_{3x}$La$_{2/3-x}$TiO$_3$~\cite{1999-Emery-JPCM-11-10401}.
Thus if a significant fraction of \elip\ forms the bound complex here, then its relaxation remains measurable even at low field.

\section{The $4c$ Site\label{app:site}}

The quadrupole splittings depend sensitively on the \elip\ site and its symmetry, and with the angular dependence of the splittings [Eqs.~\eqref{eq:nuq}, \eqref{eq:quadrupole}, and \eqref{eq:angles}] combined with calculations of the \efg\ (including lattice relaxation), one might be able to make a site assignment.
Here we set out a few properties of the most likely site ($4c$) to make some initial observations based on the site symmetry in an ideal lattice.

\begin{figure}[]
\centering
\includegraphics[width=0.45\textwidth]{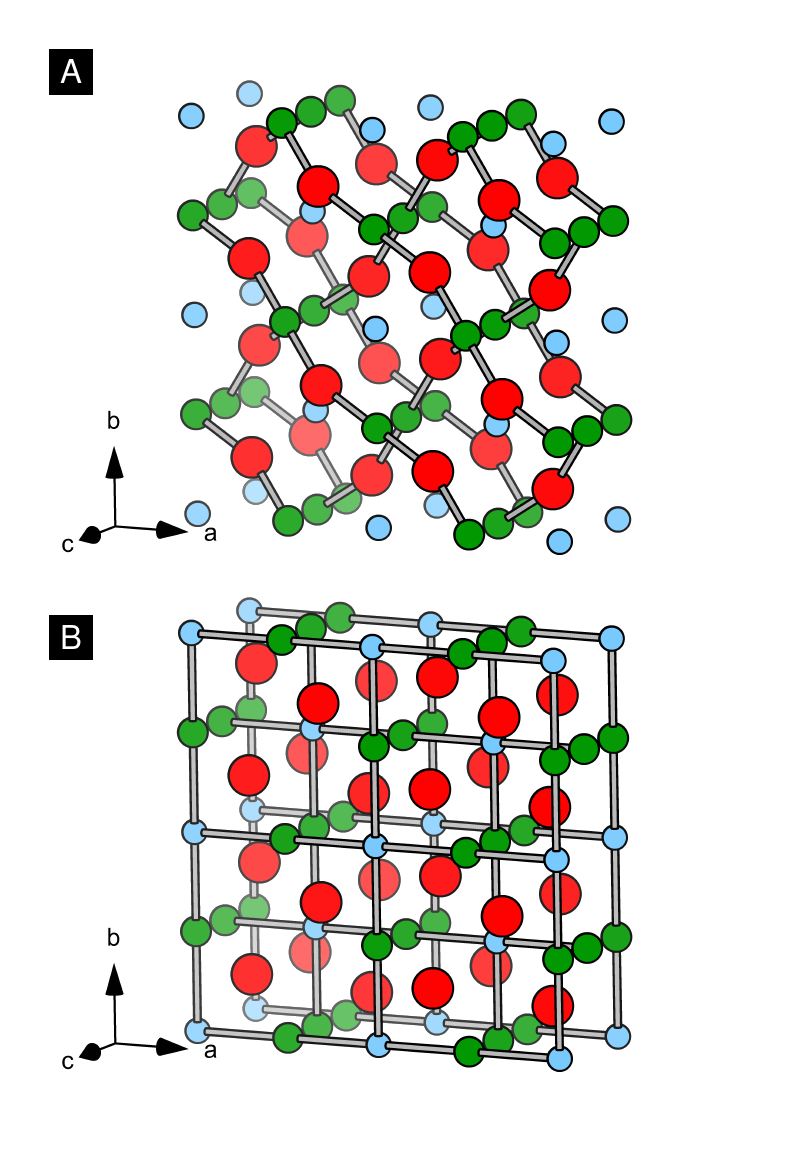}
\caption{\label{fig:coordination}
View of the near-neighbour atoms to lithium(green) in the $4c$ site, showing their \SI{90}{\degree} alternating coordination axis between neighbouring $4c$ sites along the $c$-axis channels.
The nearest-neighbour atoms are indicated here by connecting grey cylinders.
(A) The two nearest-neighbour oxygen(red) atoms.
(B) Two nearest-neighbour titanium(blue) atoms.
}
\end{figure}

The $4c$ site is coordinated by two near-neighbour and four more distant oxide ions in a shortened octahedron.
As can be seen in Fig.~\ref{fig:coordination}A, the axis of the two nearest-neighbour oxide ions alternates from one site to the next along $c$ by \SI{90}{\degree}.
Beyond the first coordination, there are two nearest neighbour Ti on opposite sides, and the $\mathrm{Ti}\mathrm{-}\mathrm{Li}\mathrm{-}\mathrm{Ti}$ direction also alternates between [100] and [010] along the channel, shown in Fig.~\ref{fig:coordination}B.
Overall, the site has $2/m$ symmetry with the two-fold axis parallel to $c$.
This symmetry is too low to yield an axial \efg, so $\eta$ is nonzero, and may even approach 1;
however, if we consider fast hopping along the $c$-axis, and the alternating character of the adjacent sites, one expects that the time-average \efg\ will become (four-fold) axisymmetric.
This should be the case at room temperature, where there is clear evidence for \lip\ diffusion, and a simple test of an axial angle dependence could confirm this.

The characteristics of this site are also important in determining properties of the \lip -polaron complex.
Here, one of the two neighbouring titanium is $3+$, rather than $4+$, which further lowers the site symmetry and alters the \efg.
Calculations suggest that the polaron is mainly mobile along the stacks of adjacent edge-sharing TiO$_6$ octahedra in the $c$ direction~\cite{2007-Deskins-PRB-75-195212, 2009-Kerisit-JPCC-113-20998, 2013-Janotti-PSSRRL-7-199, 2015-Yan-PCCP-17-29949} and it is unable to move to the other Ti neighbour on the far side of the Li site.
Notice that if the polaron does hop to the next Ti along the stack, it is not the near neighbour of the adjacent Li site, but of the second nearest Li site along the chain.
This is easily seen by the alternation of $\mathrm{Ti}\mathrm{-}\mathrm{Li}\mathrm{-}\mathrm{Ti}$ mentioned above (see Fig.~\ref{fig:coordination}B).
If the bound complex moves in tandem, with the polaron remaining in a single TiO$_6$ stack, then fast motional averaging should not result in an axisymmetric \efg, in contrast to the case of free \lip diffusion.

\begin{acknowledgments}
We thank R. Abasalti, P.-A. Amaudruz, D.J. Arseneau, S. Daviel, B. Hitti, and D. Vyas for their excellent technical support.
This work was supported by NSERC Discovery grants to R.F.K. and W.A.M.
R.M.L.M. and A.C. acknowledge the additional support of their NSERC CREATE IsoSiM Fellowships.
Crystal structure images were generated using CrystalMaker\textsuperscript{\textregistered}: a crystal and molecular structures program for Mac and Windows.
CrystalMaker Software Ltd, Oxford, England (\url{www.crystalmaker.com}).
\end{acknowledgments}

\bibliography{bnmr-rutile-v14-arxiv.bib}

\end{document}